\documentclass[conference]{IEEEtran}
\IEEEoverridecommandlockouts
\usepackage{cite}
\usepackage{amsmath,amssymb,amsfonts}
\usepackage{algorithmic}
\usepackage{graphicx}
\ifCLASSOPTIONcompsoc
    \usepackage[caption=false, font=normalsize, labelfont=sf, textfont=sf]{subfig}
\else
\usepackage[caption=false, font=footnotesize]{subfig}
\graphicspath{ {./images/} }
\usepackage{textcomp}
\usepackage{xcolor}
\def\BibTeX{{\rm B\kern-.05em{\sc i\kern-.025em b}\kern-.08em
    T\kern-.1667em\lower.7ex\hbox{E}\kern-.125emX}}
\begin{document}

\title{Against the Others! Detecting Moral Outrage in Social Media Networks}

\author{
\IEEEauthorblockN{Wienke Strathern\IEEEauthorrefmark{1}, Mirco Schoenfeld\IEEEauthorrefmark{2}, Raji Ghawi\IEEEauthorrefmark{1}, and Juergen Pfeffer\IEEEauthorrefmark{1}}
\and\and
\IEEEauthorblockA{\hspace{1cm}\IEEEauthorrefmark{1}
{Bavarian School of Public Policy}\\
\hspace{1.5cm}{Technical University of Munich, Munich, Germany} \\
\hspace{1.5cm}\{wienke.strathern, raji.ghawi, juergen.pfeffer\}@tum.de}
\and
\IEEEauthorblockA{\IEEEauthorrefmark{2}
{University of Bayreuth}\\
{Bayreuth, Germany} \\
mirco.schoenfeld@uni-bayreuth.de}
}

\maketitle

\begin{abstract}
Online firestorms on Twitter are seemingly arbitrarily occurring outrages towards people, companies, media campaigns and politicians. Moral outrages can create an excessive collective aggressiveness against one single argument, one single word, or one action of a person resulting in hateful speech. With a collective ``against the others" the negative dynamics often start. Using data from Twitter, we explored the starting points of several firestorm outbreaks. As a social media platform with hundreds of millions of users interacting in real-time on topics and events all over the world, Twitter serves as a social sensor for online discussions and is known for quick and often emotional disputes. The main question we pose in this article, is whether we can detect the outbreak of a firestorm. Given 21 online firestorms on Twitter, the key questions regarding the anomaly detection are:
1) How can we detect the changing point?
2) How can we distinguish the features that cause a moral outrage?
In this paper we examine these challenges developing a method to detect the point of change systematically spotting on linguistic cues of tweets. We are able to detect outbreaks of firestorms early and precisely only by applying linguistic cues. 
The results of our work can help detect negative dynamics and may have the potential for individuals, companies, and governments to mitigate hate in social media networks.
\end{abstract}

\begin{IEEEkeywords}
Firestorms, Twitter, Change Detection
% Social Network Analysis, 
\end{IEEEkeywords}

%%%%%%%%%%%%%%%%%%%%%%%%%%%%%%%%%%%%%%%%%%%%%%%%%%%%%%

\section{Introduction}
\label{sec:introduction}
Twitter is a social media platform with millions of users exchanging ideas about daily topics \cite{boyd2010}. Its influence on societal processes is widely discussed. Acting as social sensors for real-time discussions, users provide information about ongoing discussions for live events and media topic strategies. User interactions have provided real-time information about the success and not-success of media campaigns and public relation events. One phenomena are online firestorms, also known as electronic Word-of-Mouth. Firestorms \cite{Pfeffer2014}, thus, are not limited to failed media campaigns, but also occur in the context of verbal insult of people from politics, business and society. Negative dynamics might be very dangerous in real life and can do harm to people. A single statement or media outlet could trigger a collective brawl that seems to escalate uncontrollably until a certain point of exhaustion. 
Firestorms in Twitter have been broadly discussed in literature \cite{Pfeffer2014,Hemank2015,Rost2016,Mochalova2014,Drasch2015,Johnen2018,Stich2014}. 

\textbf{Research questions.} More generally speaking, Twitter is a source of real-time information related to discussions, topics and events that happen in society, politics, economy and media. Analyzing communication data on Twitter can help to understand the development of processes, thus, explicitly the emergence of online moral outrages, negative dynamics and collective action\cite{Crockett2017}. 
Hence, the key research questions of our study in this paper are,  "what are the major features that indicate the outbreak of a firestorm?" and "how can we detect relevant occurrence by exploring firestorm data?" Answering these questions makes researcher understand better to which extent users form an outrage collectively. Is there an "against the others"?

To answer these questions, our assumption is that a firestorm affects user positively or negatively. This leads to the question on how to approach affects in comments. We state that the words we use indicate the way we speak, hence behave and feel. Changes in the use of language in textual comments as tweets in the mention network, can give insights about changing behavior and changing perspective. It is generally accepted that in speaking we perform actions of various kinds and so we might see changes in the way a user talks and uses words\cite{Austin1975}. To understand the usage of words, we take a closer look at the tweets vocabulary from a linguistic and psychological point of view. 

\textbf{Methods.} In order to address our research question of the relationship between lexical features and firestorm participation, we use the extracted characteristics of firestorms to detect an outbreak at an early stage. 
Our approach provides a method from network analysis and text statistics by examining the dynamics of linguistic cues over time. 

To get insights into the textual and sentimental features, we evaluate the sentiment score for each tweet in comparison to the overall network and evaluate the pronouns of each tweet of a user as an indicator for changing behavior. While verbs indicate the action, pronouns indicate to whom the action is performed. Taking a closer look at the usage of pronouns,  tweets from the week before the firestorm and the week during the firestorm were analyzed.

To perform the sentiment and textual features analysis, we employ the LIWC classification tool \cite{Tausczik2009}. We focus on the positive and negative values of tweets and specific other categories of the LIWC scheme such as personal pronouns, affective processes, cognitive processes, perceptual processes and informal language gives statistical insights. Applied positivity and negativity metrics are combined with linguistic cues and compared to network effects. Lexical features have been widely explored in the psychological and computational literature. For these features we drew mainly on the LIWC lexicons of Pennebaker et al. \cite{Pennebaker2015}, a standard for social psychological analysis of lexical features. 

Our approach does not limit the analysis to just sentiments and emotional tone since it is questionable to examine the correct meaning of a tweet out of counting words. Comparing samples of the firestorm data manually to the LIWC analysis, it can be stated that many cynical statements were not captured as such. Relying simply on this valence would not give much insights about the emotional arousal of a users textual comment. 

On this account, we assume that detecting change based on sentiment analysis plus the usage of pronouns is more significant in how people connect with each other to form an outrage. Combining the automated processes that is done by the LIWCTool and looking for explicit lexical features, will help to answer the above posted questions. 

Function words are psychologically and linguistically interesting and have been studied broadly \cite{Dekker2002}. Pronouns refer to a referent, hence, tell to whom somebody is speaking \cite{Benveniste1971}. In this way, we might figure out if actors in social media networks stop talking about themselves and start talking collectively against somebody emotionally and with the words they use. The personal mode, which is the mode of dialogue, is marked by the central role of pronouns I/You. The linguist philosopher Benveniste brings into radical opposition the two first persons, necessarily posited on the basis of the I. They designate the one who speaks and imply at the same time an utterance about 'I' to the addressed object. The I-You relation founds the very exercise of speech. 
Function words are used at high rates. They require social skills to use properly. A  speaker assumes the listener knows who everyone is and a listener must know the speaker to follow the conversation. The ability to understand a simple conversation packed full of function words demands social knowledge. All function words work in this way. The ability to use them is a marker of basic social skills – and analysing how people use function words reveals a great deal about their social worlds. \cite{Pennebaker2011}

\textbf{Contributions.} A major challenge, however, is to find the signal in a firestorm that indicates anomaly. The goal is to detect sentimental and lexical changes as a signal of an underlying change in a social network.
In summary, our contributions are: 
\begin{itemize}
\item Model: We propose a novel change detection model that accounts for linguistic cues and is able to detect the outbreak of a firestorm closely and quickly.
\item Algorithm: We are able to detect firestorms on streaming Twitter data by only monitoring a couple of lexical features.
\end{itemize}

%%%======================================

\section{Related Work}
\label{sec:relwork}
In 1947 Gordon Allport and Leo Postman defined a rumor as a “proposition for belief, passed along from person to person, usually by word-of-mouth, without secure standards of evidence being presented” \cite[p~ix]{Allport1947}. While online firestorms are similar to rumors to some extent, e.g. they often rely on hearsay and uncertainty, online firestorm pose new challenges due to the speed and potential global reach of social media dynamics \cite{Pfeffer2014}. 
Why do people join online firestorms? Based on the concept of moral panics the authors argue that participation behavior is driven by a moral compass and a desire for social recognition \cite{Johnen2018}. Social norm theory refers to understand online aggression in a social-political online setting, challenging the popular assumption that online anonymity is one of the principle factors that promotes aggression\cite{Rost2016}. 

With respect to firestorms on social media, the analysis of dynamics and their early detection often involves research from the field of sentiment analysis, network analysis, and sometimes time series analysis and change point detection. 

\textbf{Sentiment analysis.} 
Approaches to the analysis of firestorms focusing on the mood of the users and their expressed sentiments unveil, for example, that in the context of online firestorms, non-anonymous individuals are more aggressive compared to anonymous individuals \cite{Rost2016}. Online firestorms are used as a topic of news coverage by journalists and explores journalists’ contribution to attempts of online scandalization. By covering the outcry, journalists elevate it onto a mainstream communication platform and support the process of scandalization. Based on a typology of online firestorms, the authors have found that the majority of cases address events of perceived discrimination and moral misconduct aiming at societal change\cite{Stich2014}. Online firestorms on social media have been studied to design an Online Firestorm Detector that includes an algorithm inspired by epidemiological surveillance systems using real-world data from a firestorm \cite{Drasch2015}.  

Sentiment analysis was applied to analyze the emotional shape of moral discussions in social networks \cite{Brady2017}. It has been argued that moral-emotional language increased diffusion more strongly. Highlighting the importance of emotion in the social transmission of moral ideas, the authors demonstrate the utility of social network methods for studying morality. A different approach is to measure emotional contagion in social media and networks by evaluating the emotional valence of content the users are exposed to before posting their own tweets \cite{Ferrara2015}. Modeling collective sentiment on Twitter gave helpful insights about the mathematical approach to sentiment dynamics \cite{Charlton2016}.

Arguing that rational and emotional styles of communication have strong influence on conversational dynamics, sentiments were the basis to measure the frequency of cognitive and emotional language on Facebook. \cite{Bail2017}. 

Instead, the analysis of linguistic patterns was used to understand affective arousal and linguist output\cite{Sharp2004}. Extracting the patterns of word choice in an online social platform reflecting on pronouns is one way to characterize how a community forms in response to adverse events such as a terrorist attack\cite{Shaikh2017}. Synchronized verbal behavior can reveal important information about social dynamics. The effectiveness of using language to predict change in social psychological factors of interest can be demonstrated nicely \cite{Gonzales2010}.
In \cite{Hemank2015}, the authors detected and described 21 online firestorms discussing their impact on the network. To advance knowledge about firestorms and the spread of rumours, we use the extracted data as starting point to follow up on the research findings. 

\textbf{Network analysis.}
Social media dynamics can be described with models and methods of social networks \cite{Wasserman:1994,Newman:2010,Hennig:2012}. Approaches mainly evaluating network dynamics are, for example, proposed by Snijders et al. Here, network dynamics were modeled as network panel data \cite{Snijders2010}. The assumption is that the observed data are discrete observations of a continuous-time Markov process on the space of all directed graphs on a given node set, in which changes in tie variables are independent conditional on the current graph. The model for tie changes is parametric and designed for applications to social network analysis, where the network dynamics can be interpreted as being generated by choices made by the social actors represented by the nodes of the graph.
This study demonstrated ways in which network structure reacts to users posting and sharing content. While examining the complete dynamics of the Twitter information network, the authors showed where users post and reshare information while creating and destroying connections. Dynamics of network structure can be characterized by steady rates of change, interrupted by sudden bursts \cite{Myers2014}. 
Network dynamics were modeled as a class of statistical models for longitudinal network data \cite{Snijders2001}. Dynamics of online firestorms where analyzed applying an agent-based computer simulation (ABS) \cite{Hauser2017}---information diffusion and opinion adoption are triggered by negative conflict messages.

\textbf{Time series analysis.}
An approach that monitors network dynamics to determine when significant change to the network structure occurs in order to more efficiently search for potential causes of change was proposed by McCulloh and Carley \cite{McCulloh2011}. The authors propose that techniques from social network analysis, combined with those from statistical process control can be used to detect when significant change occur in longitudinal network data. Statistical process control is a collection of algorithms that monitors a stochastic process over time and rapidly detects statistically significant departures from typical behavior. The approach is to mitigate negative process behavior. In order to detect change in statistical process control, periodic samples from the process were taken, calculating a statistic based on some process metric, and comparing the statistic against a decision interval. If the statistic exceeds the decision interval, the control chart is said to signal that a change may have occurred in the process.  

\textbf{Change point detection.} Detecting changes in general time series is a complex problem for itself. The most known approaches for change point detection include Binary Segmentation \cite{cpbinarysegment1,cpbinarysegment2}, Segment Neighborhood \cite{segmentneigh}, and Optimal Partitioning \cite{PELTbase}, all of which suffer from certain drawbacks when considering monitoring streaming data: Binary Segmentation is quite efficient in terms of computational complexity, i.e. $\mathcal{O}(n\log{}n)$, but it cannot guarantee to find the global minimum. Segment Neighborhood approaches suffer from computational complexity which might degenerate to $\mathcal{O}(n^3)$. Even the improvements in the form of Optimal Partitioning have a computational complexity of $\mathcal{O}(n^2)$, i.e. they are quadratic in the number of samples. 

A more recent approach was proposed by Killick et al. and it is based on the Optimal Partitioning in that it yields a guaranteed identification of the exact minimum while retaining a computational complexity that is linear in the number of samples $n$ \cite{PELTchangepoint}. Their approach is called Pruned Exact Linear Time (PELT) method and is based on a work by Jackson et al. \cite{PELTbase}. As an improvement, Killick et al. incorporated a pruning step in the dynamic program which improves the computational efficiency without impacting the exactness of the segmentation. Most importantly, their method has a linear computational complexity which renders it especially useful for applications on streaming data. 

\textbf{Mixed approaches.}
More recent approaches analyse online firestorms by analyzing both content and structural information. A text-mining study on online firestorms evaluates negative eWOM that demonstrates distinct impacts of high- and low-arousal emotions, structural tie strength, and linguistic style match (between sender and brand community) on firestorm potential \cite{Herhausen2019}. Online Firestorms were studied to develop optimized forms of counteraction, which engage individuals to act as supporters and initiate the spread of positive word-of-mouth, helping to constrain the firestorm as much as possible \cite{Mochalova2014}. 

By monitoring both linguistical and psychological features of anomaly in the mention networks of online firestorms, we also combine analysis of content with the focus on structural information. To be able to detect online firestorms quickly, we also employ a method of change point detection on time series of the extracted features.

% Related work regarding change point detection

\begin{table}[t]
\centering
\setlength\tabcolsep{4pt}
\caption{Firestorm events sorted by the number of tweets.}
\label{tab:firestorms}
\begin{tabular}{l | r r r }
Firestorm Hashtag/Mention & Tweets & Users & First Day \\
\hline
\#whyimvotingukip & 39,969 & 32,382 & 2014-05-21 \\
\#muslimrage      & 15,721 & 11,952 & 2012-09-17 \\
\#CancelColbert   & 13,277 & 10,353 & 2014-03-28 \\
\#myNYPD          & 12,762 & 10,362 & 2014-04-23 \\
\#AskThicke       & 11,763 &  9,699 & 2014-07-01 \\
@TheOnion        &  9,959 &  8,803 & 2013-02-25 \\
@KLM             &  8,716 &  8,050 & 2014-06-29 \\
\#qantas          &  8,649 &  5,405 & 2011-10-29 \\
@David\_Cameron  &  7,096 &  6,447 & 2014-03-06 \\
suey\_park       &  6,919 &  3,854 & 2014-03-28 \\
@celebboutique   &  6,679 &  6,189 & 2012-07-20 \\
@GaelGarciaB     &  6,646 &  6,234 & 2014-06-29 \\
\#NotIntendedToBeAFactualStatement & 6,261 & 4,389 & 2011-04-13 \\
\#AskJPM        & 4,321 & 3,418 & 2013-11-14 \\
@SpaghettiOs   & 2,890 & 2,704 & 2013-12-07 \\
\#McDStories    & 2,374 & 1,993 & 2012-01-24 \\
\#AskBG         & 2,221 & 1,933 & 2013-10-17 \\
\#QantasLuxury  & 2,098 & 1,658 & 2011-11-22 \\
\#VogueArticles & 1,894 & 1,819 & 2014-09-14 \\
@fafsa         & 1,828 & 1,693 & 2014-06-25 \\
@UKinUSA       &   142 &   140 & 2014-08-27 \\
\end{tabular}
\end{table}

%%%======================================
\section{Data}
\label{sec:data}

%Raji (feel free to edit)
We used the same set of firestorms as in \cite{Hemank2015}, whose data source is an archive of the Twitter decahose, a random 10\% sample of all tweets. 
This is a scaled up version of Twitter's Sample API, which gives a stream of a random 1\% sample of all tweets. Mention and re-tweet networks based on these samples can be considered as \emph{random edge sampled} networks\cite{Wagner2017} since sampling and network construction is based on Tweets that constitutes the links in the network.
% As found by Morstatter et al. [41], the Sample API (unlike the Streaming API) indeed gives an accurate representation of the relative frequencies of hashtags over time. 
% We assume that the decahose has this property as well, with the significant benefit that it gives us more statistical power to estimate the true size of smaller events.

\begin{figure}[t]
\centering
\includegraphics[width=.9\linewidth]{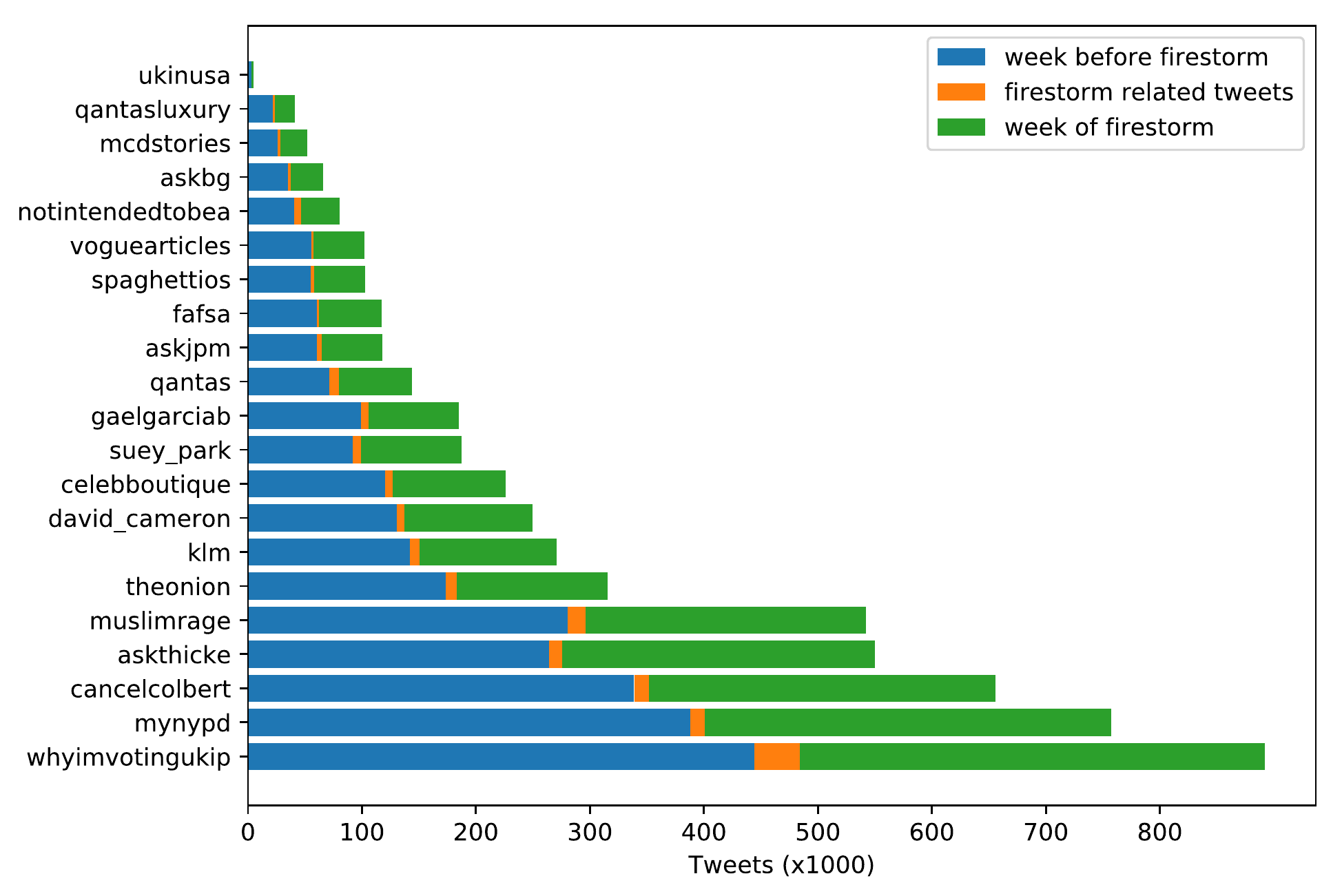}
\caption{Tweets per event}
\label{fig:events_tweets}
\end{figure}

The dataset consists of 21 firestorms with highest volume of tweets as identified in \cite{Hemank2015}.
Table \ref{tab:firestorms} shows those events along with the number of tweets, number of users, and the date of the first day of event.
The set of tweets of each firestorm covers the first week of the event.
We also augmented this dataset via including additional tweets, of the same group of users, during the same week of the event (7 days) and the week before (8 days), such that the volume of tweets is balanced between the two weeks (about 50\% each). The fraction of firestorm-related tweets is between 2\% and 8\% of the tweets of each event (Figure \ref{fig:events_tweets})---it is important to realize at this point that even for users engaging in online firestorms, this activity is a minor part of their overall activity on the platform.

For each event, we also extracted tweets metadata including timestamp, hashtags, mentions, and retweet information (user and tweet ID). 

% %Raji
% a) Dataset
% Data-wise, we have two datasets of tweets for each of the 21 firestorms:
% The first one contains the tweets within the 7 days of each event, while the second contains tweets for the same users covering the 7 days before the event. 

%This figure shows the number of tweets per hour for each event.
%frequency_h.pdf
% This figure shows a histogram of tweets per user for each event.
%user_frequency.pdf

\subsection{Mention and Retweet Networks}
To get insight on the evolution of each event, we opt to split time into units of \emph{half hours}.
This allows us to perform analysis at fine granularity. The result of this splitting is a series of about 720 time slice (since the studied time-span of an event is 15 days, this period corresponds to 720 half hours). 
At each time point we construct \emph{mention networks}, and  \emph{retweet networks} taking into account all the tweets during the last 12 hours.
This way we obtain a moving window of tweets: with window size of 24 slices at steps of half hours. 
The \emph{mention network} of each moving window contains an edge ($user_1$, $user_2$) if a tweet (among tweets under consideration) posted by $user_1$ contains a mention to $user_2$.
The \emph{retweet network} of each moving window contains an edge ($user_1$, $user_2$) if a tweet (among tweets under consideration) posted by $user_1$ is a retweet of another (original) tweet posted by $user_2$.

% The firestorm `askthicke' data is not very expressive for the analysis, that is why we removed it.

%%%======================================

\section{analysis}
\label{sec:analysis}

\subsection{Analysis of Mention Networks}

For each event, the mention networks constructed at the different time points are directed, unweighted networks.
We performed several types of social network analysis and extracted a set of metrics, including:
\begin{itemize}
\item Number of nodes $N$, and edges $E$, and density. 
\item Average out-degree (which equals avg. in-degree).
\item Maximum out-degree, and maximum in-degree.
\item Absolute and relative size of the largest connected component.
\end{itemize}

Moreover, we extracted several useful features of the corresponding moving windows, such as: number of (all) tweets, and number of mention tweets (tweets containing mentions).
We also calculated additional features, including:
\begin{itemize}
\item Ratio of mention tweets to all tweets.
\item Mention per tweet ratio: $E$ / nr. tweets.
\item Mention per `mention' user ratio:  $E$ / $N$.
\item Tweet per `mention' user ratio: nr. tweets / $N$.
\end{itemize}

\begin{figure}[b]
\centering
\includegraphics[width=\linewidth]{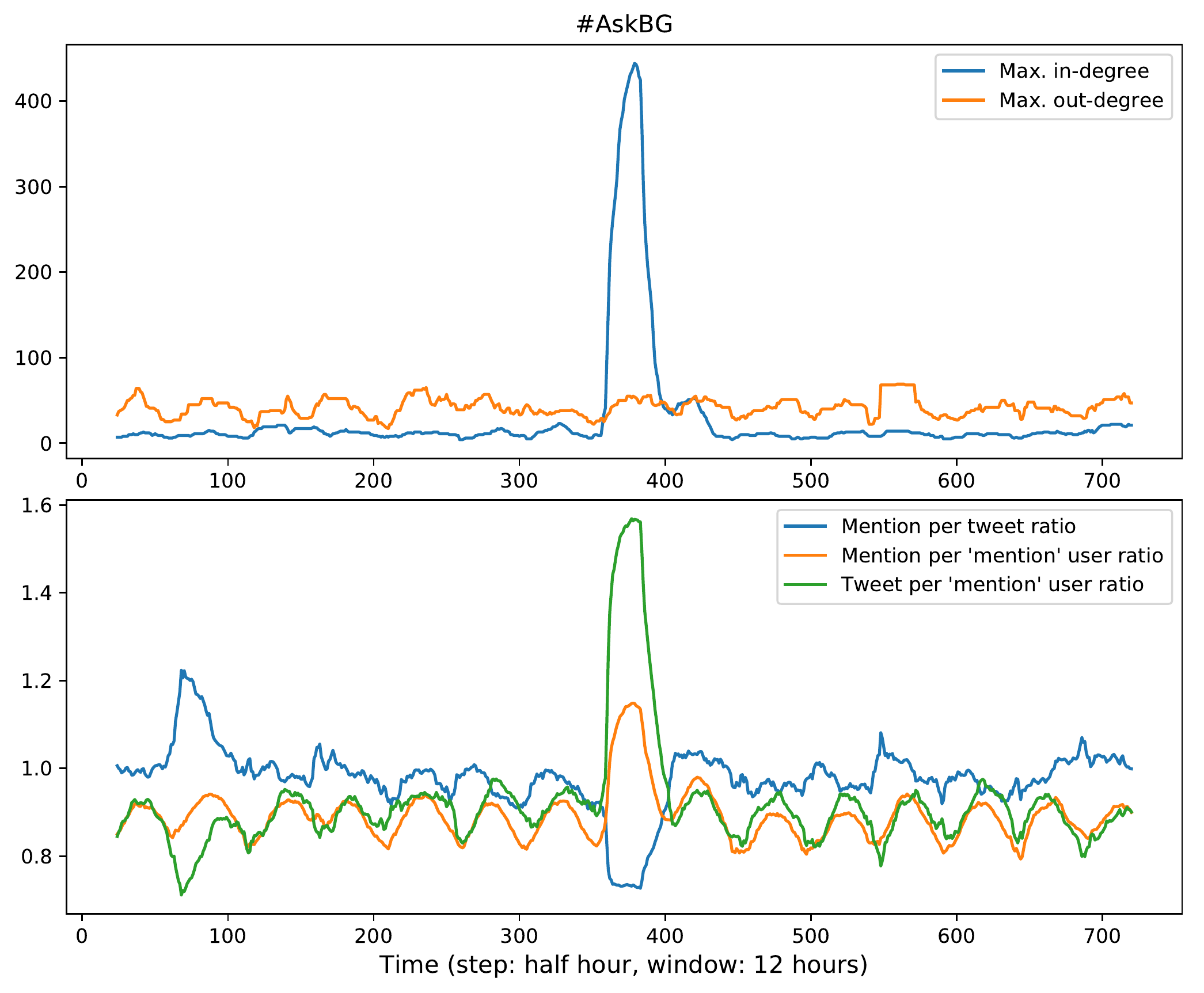}
\caption{Evolution of mention network features (\#AskBG event).}
\label{fig:muslimrage_features}
\end{figure}

Each of the aforementioned features leads to a time-series when taken over the entire time-span of the event. 
For example, Figure \ref{fig:muslimrage_features} depicts those time-series for the features of the mention networks of \#AskBG firestorm, showing how those features evolve over time.

One can clearly observe the oscillating behaviour of those features. This oscillations is due to the alternation of tweeting activity between daytime and night. 
More interesting observation is the manifest change of behavior that occurs near the middle of the time-span, which evidently signals the beginning of the firestorm event.

\begin{figure}
\centering
\includegraphics[width=.9\linewidth]{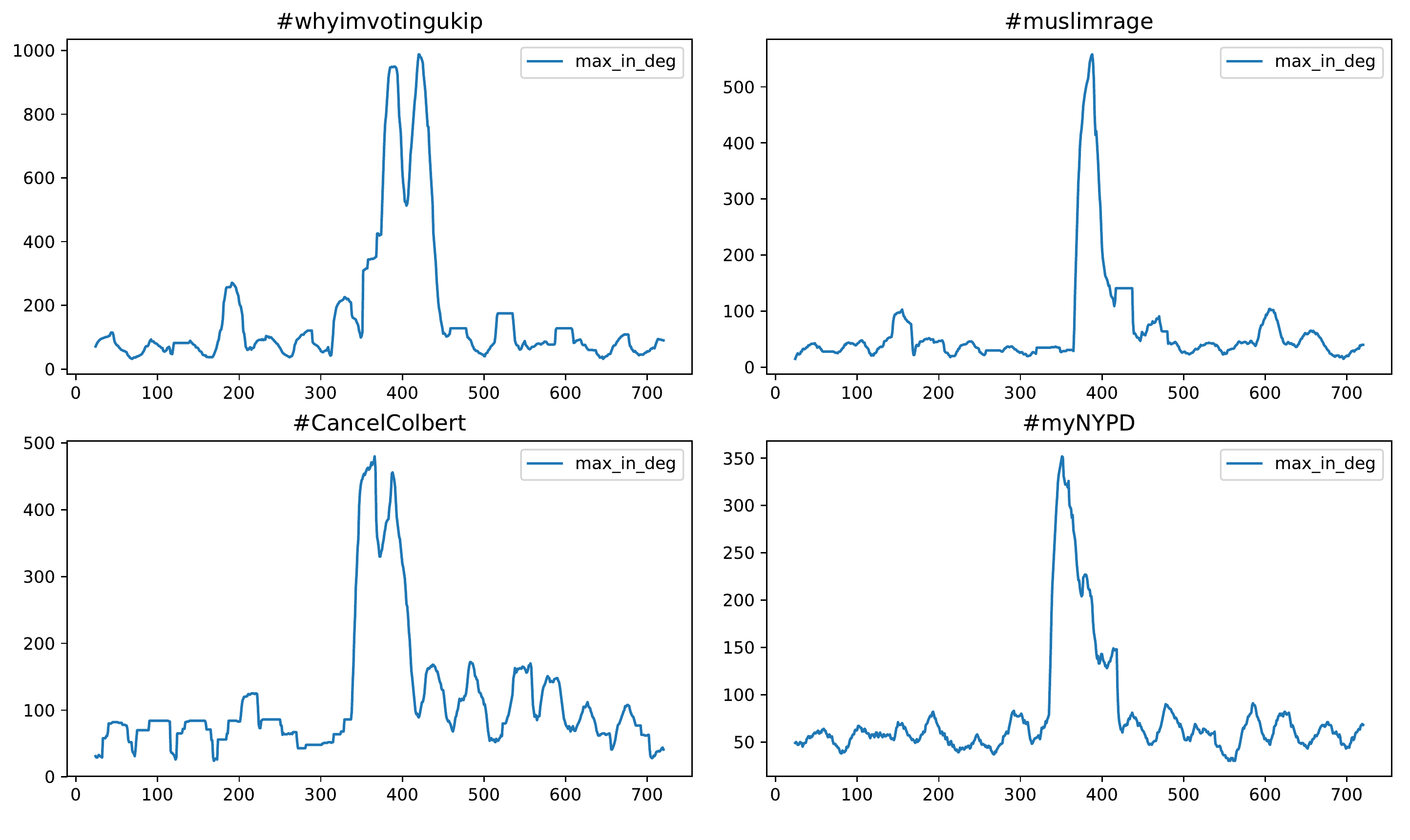}
\caption{Maximum in-degree in mention networks.}
\label{fig:max-indegree}
\end{figure}

This apparent change can be observed in most of the features for the event at hand (muslimrage).
However, not all the features are useful to detect the trigger of the firestorm in all events.
In particular, we find the maximum in-degree feature is one of the best features to detect this change.
Figure \ref{fig:max-indegree} shows the time-series of maximum in-degree for the events with largest number of tweets. Surprisingly interesting how this feature can clearly detects the start of the firestorm (in all events).
The maximum in-degree in mention networks means the highest number of mentions received by a particular user. 
Thus, the ability of this feature to detect a firestorm can be interpreted by considering that, generally speaking, a firestorm occurs when one user is being mentioned unusually high.

Thus, monitoring this feature in real-time would be certainly handy at detecting firestorms as early as possible, by signaling abnormal changes (increase) in this feature. However, the change of focus to a particular user can be the result of different (including positive) events.
A more rigours analysis of the change in behavior of such features is necessary in order to devise a formal method/algorithm of change detection as we will see in the next section.
% might help mitigate.

% feel free to edit, move or delete ;)

\subsection{Change of language}

% 1. classify texts using LIWC categories (see https://liwc.wpengine.com/wp-content/uploads/2015/11/LIWC2015_LanguageManual.pdf )
%% Personal pronouns
%% Affective processes
%% Cognitive processes
%% Perceptual processes
% 2. check for differences in these LIWC categories between firestorm and non-firestorm tweets
% 3. subset of tweets: mention networks in which people get into direct contact with each other
% Insight: 
%% some linguistic features are less prevalent during firestorms
%% especially 'i' pronouns are used a lot less. this is even significant in nearly all (20/21) firestorms
%% 'posemo' is significantly lower in firestorm tweets.
%% these observations are even stronger in mention networks

The first step was to uncover the linguistic peculiarities of firestorms. We classified all tweets using the LIWC classification scheme \cite{Pennebaker2015} and compared between firestorm tweets and non-firestorm tweets. The comparisons refer to the following categories:
\begin{itemize}
    \item Personal Pronouns,
    \item Affective Processes,
    \item Cognitive Processes, 
    \item Perceptual Processes, and
    \item Informal Language
\end{itemize}
These categories each contain several subcategories that can be subsumed under the category names. The category of personal pronouns, for example, contains several subcategories referring to personal pronouns in numerous forms. One of these subcategories `I', for example, includes -- besides the pronoun `I' -- `me', `mine', `my', and special forms such as `idk' (which means ``I don't know"). It is noteworthy that entries of subcategories may overlap, e.g. `idk' is part of `Netspeak', `Filler', `Personal Pronouns', and `I'. For such cases, and to account for imbalance between subcategories, the LIWC classification scheme assigns individual weights to matches of subcategories.

% was für werte ermittelt liwc hier?

To compare firestorm tweets and non-firestorm tweets, the single subcategories were sufficient to gain detailed insights. This allows us to compare the firestorms regarding the subcategories and to identify subcategories which distinguish the firestorm tweets more clearly from tweets from non-firestorm periods. For each individual subcategory, we obtain the mean value of the respective LIWC values for the firestorm tweets and the non-firestorm tweets.

\begin{figure}[t]
\centering
\includegraphics[width=.9\linewidth]{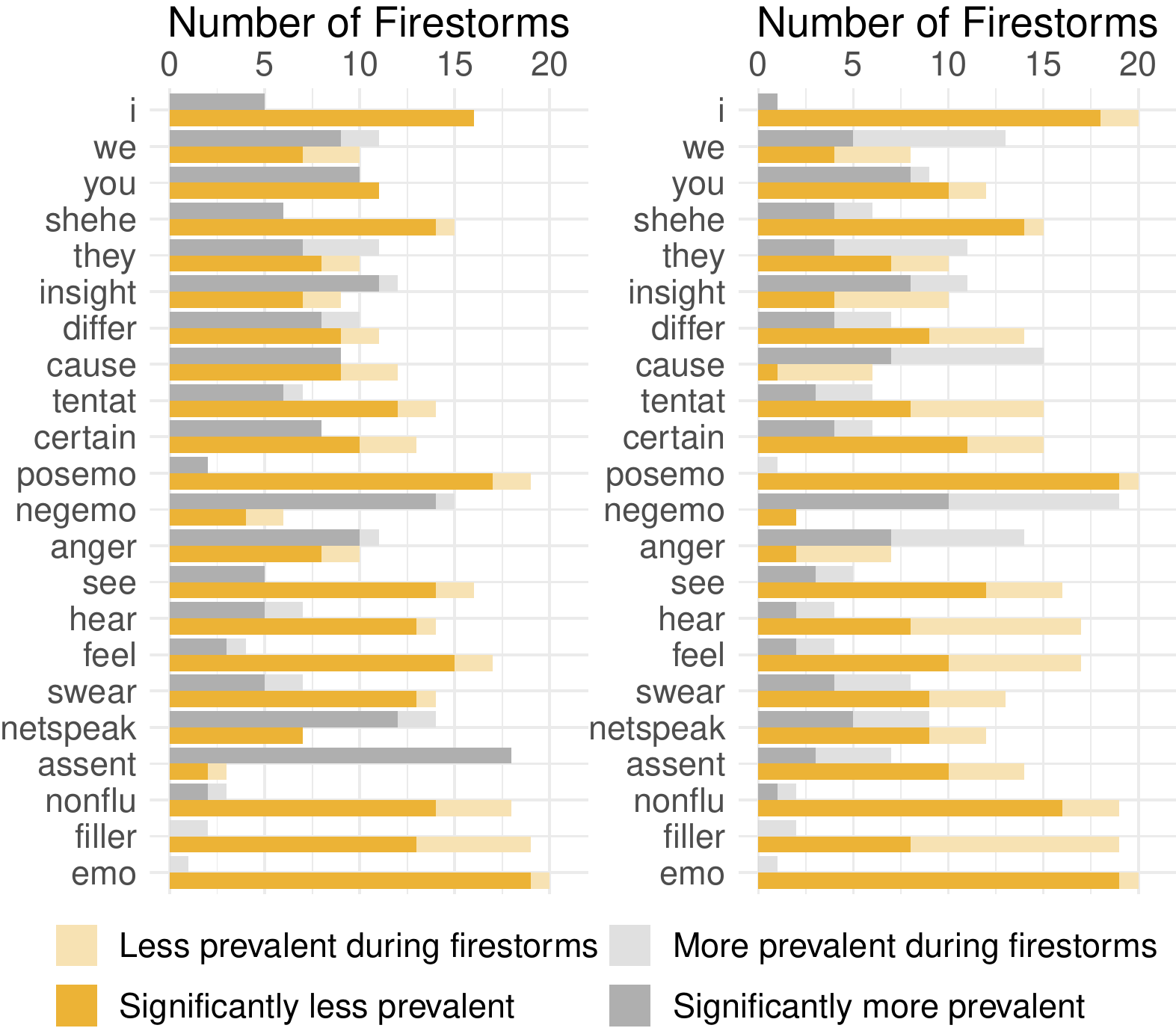}

\subfloat[All Tweets\label{fig:compare_liwc_alltweets}]{%
\hspace{.45\linewidth}
}
%\hspace{.3\linewidth}
\subfloat[Mention Networks\label{fig:compare_liwc_mentionnetworks}]{%
\hspace{.45\linewidth}
}
\caption{Comparison of linguistic features between firestorm tweets and non-firestorm tweets}
\label{fig:compare_liwc}
\end{figure}

In Figure \ref{fig:compare_liwc} the comparisons between firestorm tweets and non-firestorm tweets are shown with regard to the individual subcategories. The firestorm-tweets were compared with tweets from the week immediately before the firestorm. We conducted the comparison to the non-firestorm tweets of the week of the firestorm, but the results look very similar. Hence, we are reporting only the former comparison here.

As described in Section \ref{sec:data}, the comparison tweets contained only the users who have contributed to the firestorm. Figure \ref{fig:compare_liwc_alltweets} refers to all tweets, while for Figure \ref{fig:compare_liwc_mentionnetworks} only tweets from the @mention network were considered. In both cases every subcategory was examined separately for all 21 firestorms. 

The grey bars in the graphic represent the number of firestorms in which words from the respective category occurred more frequently during the firestorms. The orange bars visualize the number of firestorms in which the same words occurred less frequently during the firestorms. The sum of the orange and grey bars is therefore always 21. 

The light areas of the bars indicate that the mean values were different from each other. The strongly colored areas of the bars indicate that these differences were significant in terms of t-tests with $p<0.01$. For category `I' in Figure \ref{fig:compare_liwc_alltweets}, this means that in 5 Firestorms people used words of this category significantly more often, while in 16 Firestorms these words were used significantly less. Words of the same category were used less in 20 Firestorms considering the mention networks alone as depicted in Figure \ref{fig:compare_liwc_mentionnetworks}. In 19 of these Firestorms, the differences were also significant.

In addition to the category `I', the categories `posemo' and `assent' should also be highlighted. Words representing positive emotions like `love', `nice', `sweet' -- the `posemo' category -- are used clearly (and often significantly) less in almost all firestorms: positive emotions were less present in 19 out of 21 firestorms. In 17 out of 19 firestorms the differences were significant. This effect even increases when looking at @mention networks. 

For the third remarkable category `assent', which contains words like `agree', `OK', `yes', this effect is reversed for all tweets -- words in this category are used significantly more often during almost all firestorms (18 out of 21). When looking at the @mention networks, however, this feature lacks accuracy. The differences are significant only in 13 firestorms.

Finally, we constructed our own category 'emo' by calculating the difference between positive and negative sentiments in tweets. Thus, weights of this category can be negative and should describe the overall sentiment of a tweet. This is a quite common procedure in dealing with sentiment information. It can be seen how this category has even more pronounced differences between firestorm-tweets and non-firestorm tweets. There are 19 firestorms in which the 'emo' values were significantly lower during a firestorm. At the same time, there was only one firestorm with higher values of 'emo' but these differences were not significant.

Checking if the differences remain visible decomposing the @mention networks into components comparing tweets inside the largest component to tweets outside that component, we see no effect. There are only a few firestorms, in which the use of `we' is significantly larger inside the largest component of the mention network.

\subsection{Change point detection}
The goal is to identify a firestorm at an early stage with the help of linguistic features. For this purpose, we first identify the most suitable of the above mentioned characteristics. For the detection of change points, we use an efficient method that is suitable for being applied to streaming data and that was proposed by Killick et al. \cite{PELTchangepoint}. 

\begin{figure}[b]
\includegraphics[width=.82\linewidth]{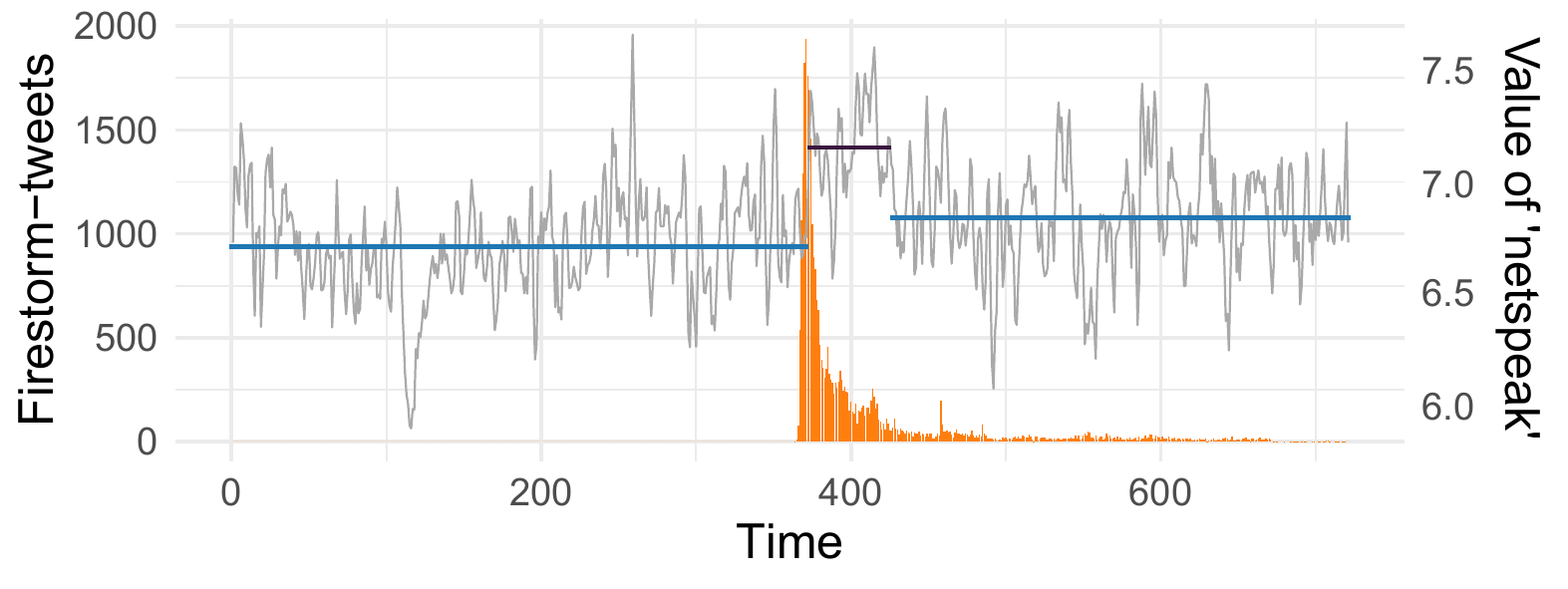}
\centering
\caption{Example change point detection conducted on a linguistic category. The number of Firestorm tweets is depicted in the background.}
\label{fig:chgpnt_example}
\end{figure}

To detect the change points, individual time series of the linguistic features have to be constructed. For this purpose, we first split the timeline of each of the firestorm data sets into buckets of half hours and assign tweets to buckets based on their timestamp. When constructing a time series of linguistic characteristics, we describe a bucket of tweets by the mean value of the corresponding LIWC values.

Figure \ref{fig:chgpnt_example} depicts an example of detecting the change points in the time series of the 'netspeak' category. The number of firestorm tweets per half hour buckets is shown in the background. The figure visualizes nicely how the main part of the firestorm is covered well by an interval between two change points. 

To detect the change points in streaming data, we simulate the arrival of new tweets every half hour. Being able to decide at any time $t$ whether a change has occurred, we use historical data from the past 24 hours. The time series applied for change point detection thus consists of 49 values for the interval $[t-48:t]$ each representing the mean of the LIWC values of the respective tweets. By doing so, we create separate time series for each of the subcategories mentioned above. We do not apply any smoothing to the time series. 

\begin{figure*}[ht]
\centering
\includegraphics[width=\linewidth]{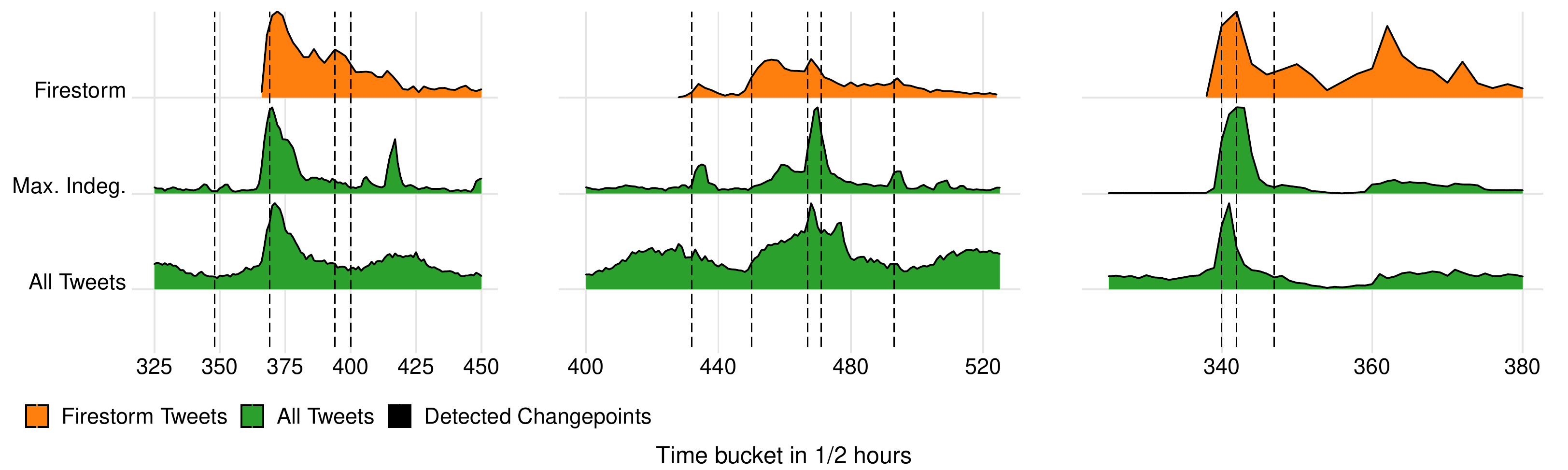}\\
\subfloat[\#muslimrage\label{fig:cps_mrage}]{%
\hspace{.31\linewidth}
}
\subfloat[\#AskThicke\label{fig:cps_askth}]{%
\hspace{.31\linewidth}
}
\subfloat[@fafsa\label{fig:cps_fafsa}]{%
\hspace{.31\linewidth}
}
\caption{Comparison of change points detected based on LIWC features with the relevant periods in time highlighted.}
\label{fig:compare_storms}
\end{figure*}

However, it should be noted that there are only 49 floating point numbers for each category. In addition, the amount of memory required is constant: Unlike observing hashtags or @mentions, for example, the words searched for in the individual tweets are defined beforehand. In live operation, no additional memory has to be reserved for newly arriving hashtags or @mentions. Ultimately, only up to half an hour of live data is required to continue the history. 

Decisive for successful change point detection is the choice of the penalty parameter. We select this parameter using  elbow criterion. Therefore, we iterate penalty parameters from 2 to 10 and obtain the number of identified change points. From this data we determine the optimal penalty parameter as the configuration with the maximum absolute second derivative using the approximation of the second derivative of a point $x_i$ as $x_{i+1} + x_{i-1} - 2x_i$. Choosing a higher penalty value results in fewer change points detected and vice versa. 

We were able to detect the start of the firestorms in $-0.55 \mp 2.34$ hours. We have defined the start time as the first interval of half an hour at which the hashtag or @mention of the firestorm was the most frequent hashtag or @mention in the data set. Due to the focused data collection process, the set contains little hashtags or @mentions that were used frequently. Hence, this definition of a starting point of firestorms is quite sensitive, i.e. a low number of tweets suffices to boost the relevant hashtag or @mention.

Figure \ref{fig:compare_storms} depicts three exemplary firestorms and the change points determined in each case. The orange curve describes the accumulation of the specified hashtags, or @mentions. Two lower curves show the course of the maximum in-degree in the @mention network and the absolute number of all tweets (firestorm tweets and non-firestorm tweets) in the respective data set for comparison. The beginning of the firestorm was identified quite reliably. 

An average value indicates the closest change point to starting time that lies about half an hour before the starting time which is one interval in our discretized timeline. With the extracted linguistic indicators, we are therefore able to detect a change in the linguistic behaviour of the users closely. 

It is important to know when this change point was detected. For this purpose, we have determined the time $t$, at which the minimum deviation from the start time was measured. It takes two intervals of half an hour until the beginning of the firestorm is noticed. Hence, we are able to detect a change in the linguistic behaviour of the users quickly. 

In a next step, we explored another change near the peak of a firestorm. Determining this peak was done in two ways. 

The first definition picks up on Figure \ref{fig:max-indegree} and describes the half-hour interval in which the highest in-degree in the @mention network was measured. We were able to approximate this peak of the network dynamics with an average of $+1.19 \mp 2.51$ hours. Means that the change point closest to the peak is on average shortly after this peak. 

The second, natural definition of the peak is the interval of half an hour in which most firestorm-related tweets were recorded, i.e. tweets which contain either the corresponding hashtag or the @mention. We were able to approximate this peak of the tweet accumulations with an average of $+0.14 \mp 1.30$ hours. The closest determined change point is shortly before the peak of the tweet accumulations. With the help of the linguistic indicators, we were able to determine another change point just before the peak of the firestorm. 

\begin{figure}[b]
\includegraphics[width=.8\linewidth]{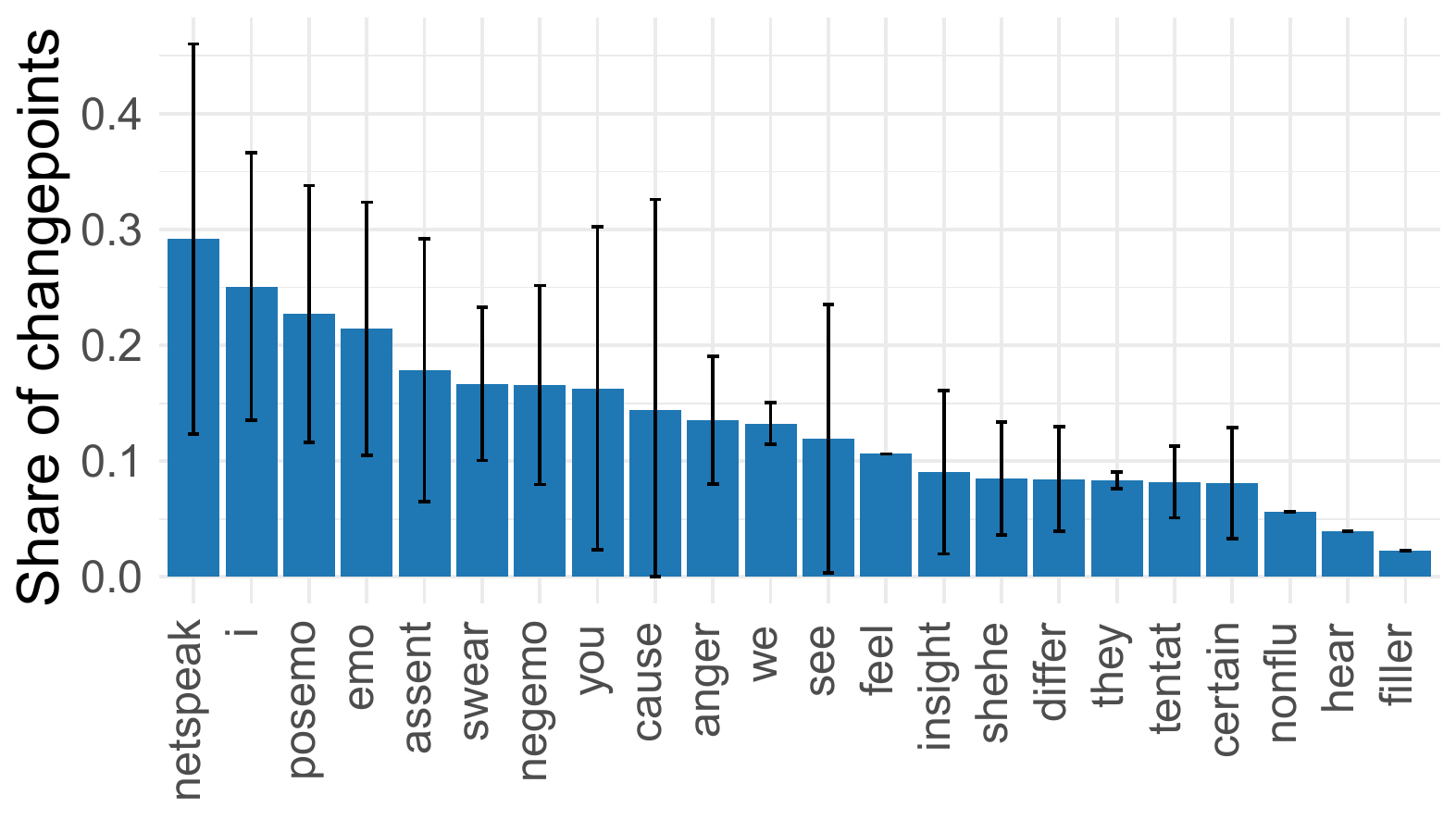}
\centering
\caption{How often were predictors relevant for the detection of a relevant changepoint}
\label{fig:chgpnt_predictors}
\end{figure}

With respect to the identified differences in language use that were discussed in relation to Figure \ref{fig:compare_liwc}, we further evaluated how many change points were identified on the timelines of the linguistic categories. Figure \ref{fig:chgpnt_predictors} depicts how often a change point could be detected from the timeline of an individual characteristic. The top three categories were 'netspeak', 'I', and 'posemo' which corresponds to the insights from Figure \ref{fig:compare_liwc} -- these were the categories with the most significant differences just after our own category 'emo'. 

From a network perspective, a firestorm occurs when one user is being mentioned unusually high -- focusing on a Twitter handle or a hashtag. The maximum in-degree in @mention networks is significantly deviating from comparable time periods. 
By evaluating lexical cues from the Tweet comments, we evaluated collective behavior manifesting in individual choices of words. During firestorms, users talk significantly less about themselves compared to non-firestorm periods. Simultaneously, the positivity in firestorms tweets vanishes and negativity rises. The extracted lexical features were applicable to streaming data. Using lexical features to monitor change in behavior has the advantage of constant memory requirements. By applying a straightforward change point detection, we were able to detect the starting point of the firestorms closely and quickly. We further provide insight into which linguistic categories proved to be useful for this task.

%%%======================================

%\section{Discussion}
%\label{sec:discussion}
%\input{sections/discussion}

%%%======================================

\section{Discussion \& Conclusion}
\label{sec:conclusion}
According to our posed questions, we conclude the following: Our research questions tackled the detection of anomaly in a network by exploring major features that indicate the outbreak of a firestorm, hence, the starting point of possible collective changing behavior. 

On this account, combining sentiment analysis and text statistics to explore firestorm data can reveal in how people connect with each other to form an outrage. 
The usage of vocabulary changes at a certain point when every single user stops commenting with the I-perspective and starts commenting on others. 
As mentioned, pronouns refer to a referent. If the 'I' diminishes, the focus changes significantly. All of a sudden people stop talking collectively about themselves positively and collectively more negatively - against the others.

Our model picks up these features and is able to detect the starting point of outrages giving insights about the changing user perspective to "against the others". 
Further research questions regarding spreading of rumours and moral outrages might be:

\begin{itemize}
\item  What causes evolving collective emotionality?
\item  Why does a community or society may at times come together and simultaneously communicate the same thought and participate in the same action?
\end{itemize}

A better knowledge of individual motivations and collective action can help to better understand and detect online firestorms.

%%%======================================

\section{Acknowledgements}
\label{sec:acknowledgements}
%Short:The author(s) gratefully acknowledge the financial support from the Technical University of Munich - Institute for Ethics in Artificial Intelligence (IEAI).
 
The author(s) gratefully acknowledge the financial support from the Technical University of Munich - Institute for Ethics in Artificial Intelligence (IEAI). Any opinions, findings and conclusions or recommendations expressed in this material are those of the author(s) and do not necessarily reflect the views of the IEAI or its partners.

%%%%%%%%%%%%%%%%%%%%%%%%%%%%%%%%%%%%%%%%%%%%%%%%%%%%%%

\bibliographystyle{IEEEtran}
\bibliography{IEEEexample}

\end{document}